\documentclass[epjCONF]{svjour}
\usepackage{graphics}
\usepackage[varg]{txfonts} % Times fonts
\usepackage[latin1]{inputenc}
\session-title{Ageing low mass stars: from red giants to white dwarfs}
\begin{document}
\title{Internal rotation of red giants by asteroseismology}
\author{M. P. Di Mauro\inst{1}\fnmsep\thanks{\email{maria.dimauro@inaf.it}} \and D. Cardini\inst{1} \and R. Ventura\inst{2} \and D. Stello\inst{3} \and P. G. Beck\inst{4} \and G. Davies\inst{5}\and Y.~Elsworth\inst{6} \and R. A. Garc\'ia\inst{5} \and S. Hekker\inst{7} \and B. Mosser\inst{8} \and J. Christensen-Dalsgaard\inst{9}  \and S.~Bloemen\inst{4} \and G. Catanzaro\inst{2} \and K. De Smedt\inst{4} \and A. Tkachenko\inst{4}}
\institute{INAF-IAPS Roma, Italy \and INAF- Osservatorio Astrofisico di Catania, Italy 
\and Sydney Institute for Astronomy, School of Physics, University of Sydney, Australia 
\and Instituut voor Sterrenkunde, Katholieke Universiteit Leuven, Belgium 
\and AIM, CEA/DSM-CNRS-Universit\'e Paris Diderot, IRFU/Sap., Certre de Saclay, France 
\and School of Physics and Astronomy, University of Birmingam, UK 
\and Astronomical Institute Anton Pannekoek, University of Amsterdam, The Netherlands 
\and LESLIA, CNRS, Universit\'e Pierre et Marie Curie, Universit\'e Denis Diderot, Observatoire de Paris, Meudon Cedex, France 
\and Department of Physics and Astronomy, Aarhus University, Denmark}
\abstract{
We present an asteroseismic approach to study the dynamics of the stellar
interior in red-giant stars by asteroseismic inversion of the splittings induced
by the stellar rotation on the oscillation frequencies. We show preliminary results
obtained for the red giant KIC4448777 observed by the space mission {\it Kepler}.
} %end of abstract
\maketitle
%
%\section{Introduction}
%\label{intro}
%Your text comes here. Separate text sections with

\section{Asteroseismic data}

\begin{figure}
{\vbox{
\hspace{-0.5cm}
\resizebox{0.65\columnwidth}{!} {
  \includegraphics{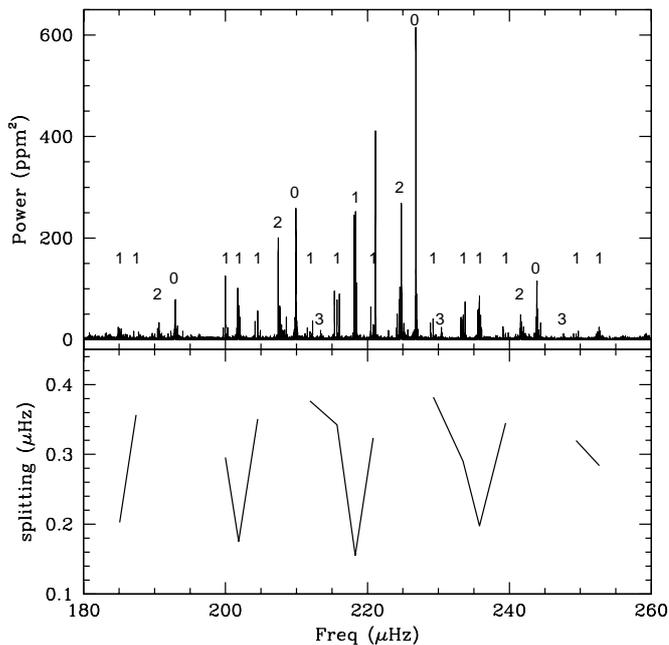}}}
\vspace{-5cm}
\hfill\parbox[b]{4.6cm}{
\caption{The upper panel shows the oscillation spectrum of KIC4448777. 
The harmonic degree of the observed modes
($l$=0,1,2,3) are indicated. Multiplets due to rotation are
visible for $l$=1. The lower panel shows the values of the
observed rotational splitting for individual $l$=1 modes.}}}
\vspace{1cm}
\label{fig:1}      
\end{figure}

The red giant KIC4448777 has been continuously observed by the {\it Kepler} satellite for 670 
days in long-cadence mode (integration time of 30 min). 
The Fourier analysis of the long time series has shown a clear power excess between 
180--260~$\mu$Hz (Fig.~\ref{fig:1}) and allowed us to identify
58 individual modes characterized by a mean large frequency separation
$\Delta\nu=16.96\pm0.03~\mu$Hz, a true period spacing $\Delta{\rm P}=90\pm3$~s \cite{RefBe} and a frequency 
of the maximum amplitude of the smoothed excess power of $\nu_{\mathrm {max}}=219.75\pm1.23~\mu$Hz. 
The observed modes are $l$=0 pure acoustic modes, and $l$=1, $l$=2 and $l$=3 modes with 
mixed gravity--pressure character.

\label{sec:1}
\section{Rotational splittings}

The stellar rotation breaks the spherical symmetry of the structure of the star and 
splits the frequencies of normal modes in (2$l$+1) components. Fig.~\ref{fig:1} shows 
that the spectrum of KIC4448777 is characterized by the presence of 15 rotational 
splittings for $l$=1. As it has been noticed by \cite{RefB}, the 
observed rotational splittings are not constant for consecutive dipole modes (see lower 
panel of Fig.~\ref{fig:1}): splittings are larger for modes with larger gravity
component which sound better the core. This indicates that the core of this star is 
rotating faster than the upper layers. In order to quantify the internal rotation it is 
possible to invert the following equation, obtained by applying a standard perturbation 
theory to the eigenfrequencies, relating the splittings 
$\delta\nu_{n,l}$ to the internal rotation $\Omega$(r):

\begin{equation}
\delta\nu_{n,l}=\int_{0}^{R} K_{n,l}(r)\frac{\Omega(r)}{2\pi}dr
\end{equation}

\noindent where $K_{n,l}(r)$ are the mode kernel functions calculated on the unperturbed
eigenfunctions for the modes $(n,l)$ of the 'best' model of the star and $R$ is the 
photospheric stellar radius.

       \begin{table*}
{\hbox{
\parbox[c]{7cm}{
\caption{Atmospheric parameters.}
\label{tab:1}  
\hspace{1.7cm}
    \begin{tabular}{|l|c|}
\hline\noalign{\smallskip}
$M_v$       & 11.56\\
\hline\noalign{\smallskip}
$T_eff$ (K) & 4750$\pm$ 250 \\
\hline\noalign{\smallskip}
log g (dex) & 3.5 $\pm$ 0.5 \\
\hline\noalign{\smallskip}
$[Fe/H] $   & 0.23$\pm$ 0.12\\
\hline\noalign{\smallskip}
$v \sin \rm{i} $ (km/s)& $<$ 5\\
\hline\noalign{\smallskip}
    \end{tabular}}
\hfill\parbox[c]{7cm}{
\vspace{-0.3cm}
\caption{Parameters of the best fitting models.}
\label{tab:2}
\hspace{1.3cm}
    \begin{tabular}{|l|c|c|}
\hline\noalign{\smallskip}
            & Model 1& Model 2\\
\hline\noalign{\smallskip}
$M/M_\odot$ & 1.02  &1.13\\
\hline\noalign{\smallskip}
$T_{eff}$ (K) & 4800  &4735 \\
\hline\noalign{\smallskip}
log g (dex) & 3.26  &3.27\\
\hline\noalign{\smallskip}
$R/R_\odot$ & 3.94  &4.08\\
\hline\noalign{\smallskip}
$L/L_\odot$ & 7.39  & 7.50\\
\hline\noalign{\smallskip}
$(Z/X)_i $ & 0.022 &0.032\\
\hline\noalign{\smallskip}
    \end{tabular}}
}}
        \end{table*}
\label{sec:2}

\section{Results and conclusion}
\begin{figure}
{\vbox{
\resizebox{0.68\columnwidth}{!} {
\includegraphics{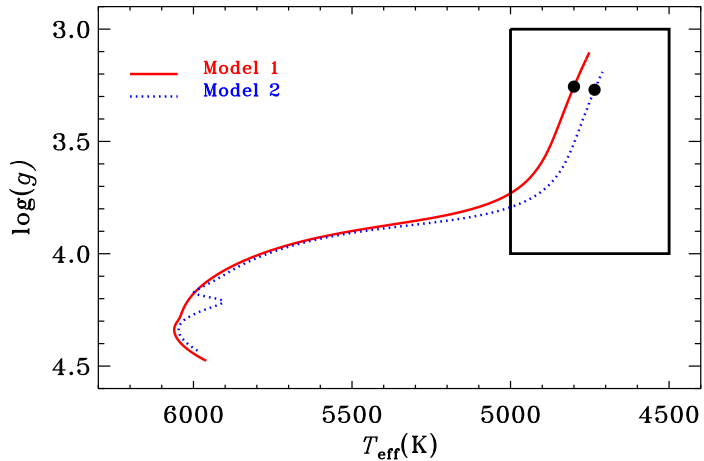}}}
\vspace{-3cm}
\hfill\parbox[b]{4cm}{\caption{Evolutionary tracks plotted in a H-R
diagram. Black dots indicate two models
which best reproduce the observations.}}}
\vspace{1.5cm}
\end{figure}
The theoretical structure models which better reproduce the identified pulsational 
frequencies have been calculated by using the ASTEC evolution code \cite{RefC1} 
assuming the basic atmospheric parameters given in the Table~\ref{tab:1}. These 
parameters have been obtained by the analysis of the spectra taken with the Hermes 
spectograph \cite{RefHe} mounted to the 1.2 m Mercator telescope. 
Adiabatic oscillation frequencies were calculated by using the ADIPLS code \cite{RefC2} 
and corrected for the surface effect by using the relation proposed by \cite{RefK}. The mass, the effective temperature, the gravity, the surface radius,
the luminosity and the initial metallicity of the two models which
best fit the observations are given in Table~\ref{tab:2}.
These values indicate that this star is in the hydrogen-shell burning phase 
(see Fig.~2), as predictable from the period spacing.

\begin{figure}
\hspace{-0.8cm}
\resizebox{1.1\columnwidth}{!} {
\includegraphics{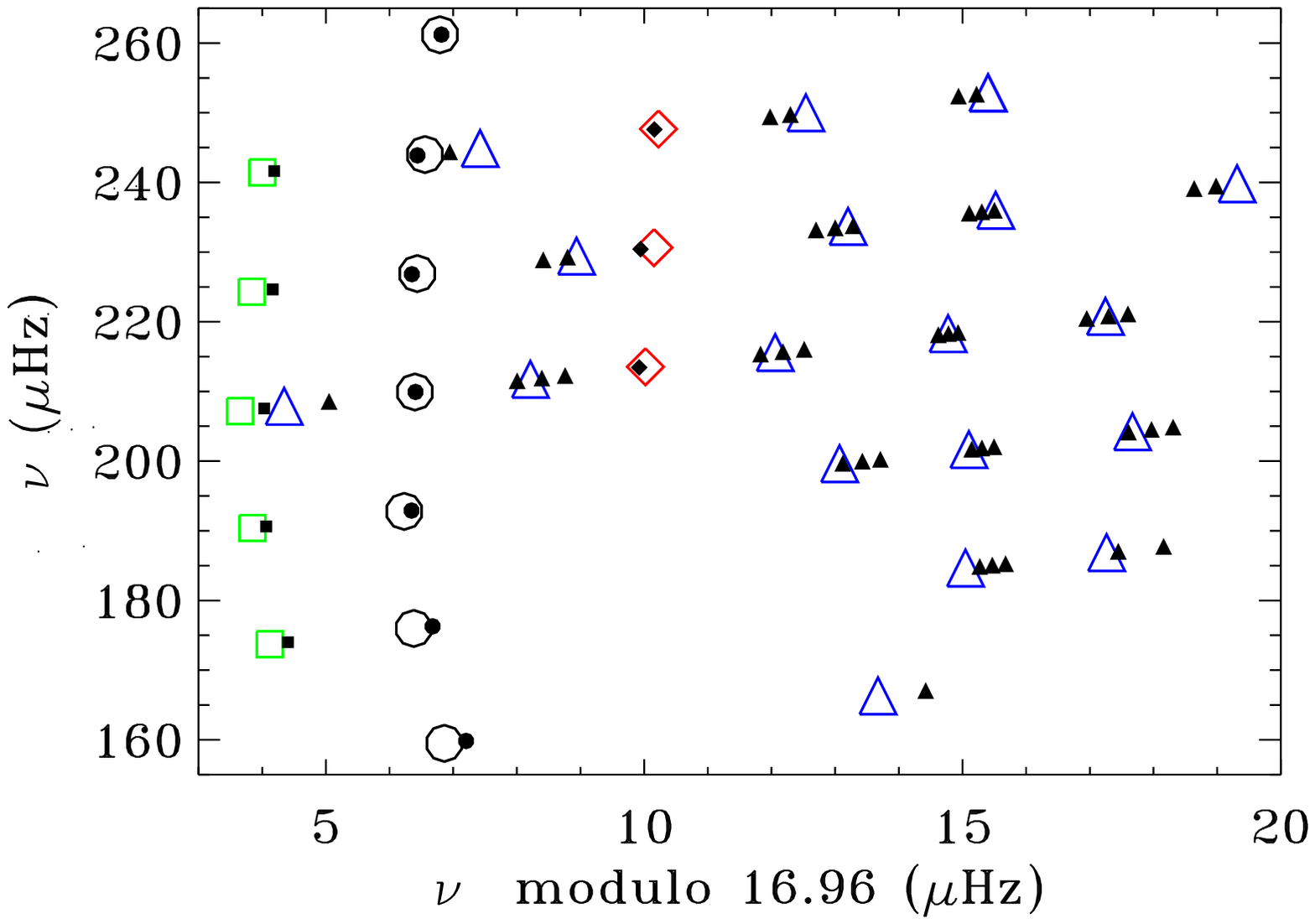}
\hspace{-1.5cm}
\includegraphics{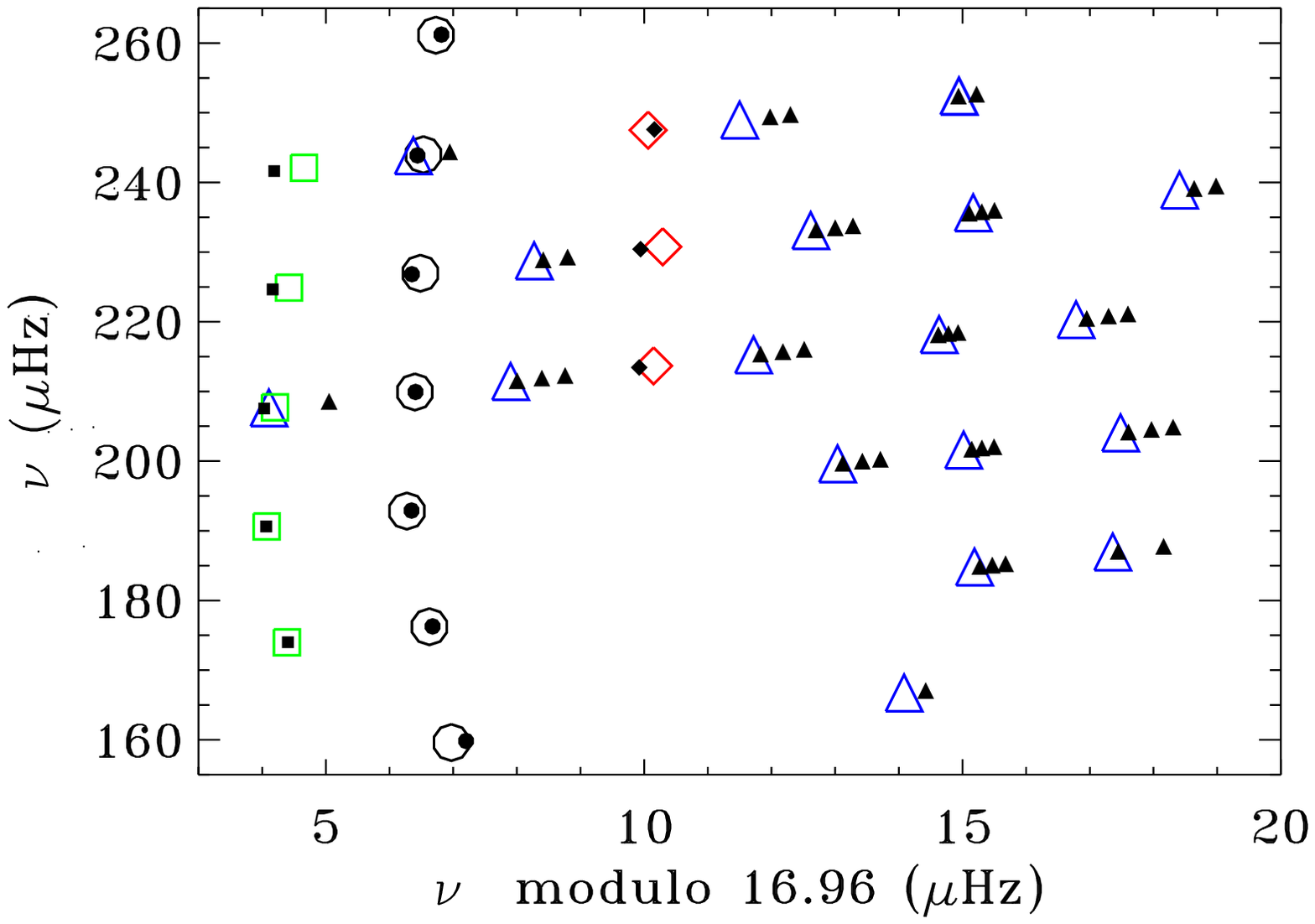}}
\caption{Echelle diagrams for Model~1 (on the left) and Model~2 (on the right) of Table~\ref{tab:2}. The
filled symbols show the observed frequencies. The
open symbols show the computed frequencies. Circles are used for modes with $l=0$, triangles for $l=1$, squares for $l=2$, diamonds for $l=3$.}
\label{fig:3}
\end{figure}

Figure~\ref{fig:3} shows the
echelle diagram obtained for one of the best fitting models. We plan to invert Eq.~1 by using both the
OLA (Optimally localized Averaging) and the SOLA (Subtractive Optimally Localized
Averaging) techniques which were successfully applied to the Sun (see e.g. \cite{RefP}; 
\cite{RefD1}). These allow to estimate a localized weighted average 
of $\Omega(r)$ making attempts to fit the
averaging kernels to a function of fixed width and centered at a chosen value of radius. Results will give
us quantitative information on the differential rotation of the interior of the examined star.
\label{sec:3}

\end{document}